\documentclass{PoS}
\usepackage{epsfig,cite}

\title{
\vspace*{-2.3cm}
\begin{minipage}{\textwidth}
{\normalfont\small LTH 968, DESY 12-237, LPN12-133, SFB/CPP-12-97
\hspace{\fill} December 2012}\\
\end{minipage}\\[35pt] 
       Progress on double-logarithmic large$\:\!$-x and small$\:\!$-x
       resummations for (semi-)$\,$inclusive hard processes}

\ShortTitle{Progress on double-logarithmic large-$x$ and small-$x$
            resummations}

\author{\speaker{A.~Vogt},~
        C.H. Kom,~
        N.A. Lo Presti\thanks{$\,$Present address: 
        Institut de Physique Th\'eorique, CEA-Saclay,
        F-91191, Gif-sur-Yvette cedex, France},~
        G. Soar\thanks{Address until September 2011}\\
        \mbox{Department of Mathematical Sciences, University of Liverpool
        Liverpool L69 3BX, UK}\\
        E-mail: \email{Andreas.Vogt@liv.ac.uk}}

\author{A.A. Almasy,~ S. Moch\thanks{Present address: 
        II.~Institut f\"ur Theoretische Physik, Universit\"at Hamburg, 
        D-22761 Hamburg, Germany}\\
        Deutsches Elektronensynchrotron DESY, 
        Platanenallee 6, D--15738 Zeuthen, Germany}
        
\author{J.A.M. Vermaseren\\
        NIKHEF Theory Group, Science Park 105, 1098 XG Amsterdam, 
        The Netherlands}
       
\author{K. Yeats\\
        Department of Mathematics, Simon Fraser University, 
        Burnaby, BC, V5A 1S6, Canada\\ \\ }

\abstract{Over the past few years considerable progress has been made on the
resummation of double-logarithmically enhanced threshold (large-$x$) and 
high-energy (small-$x$) higher-order contributions to the splitting functions 
for parton and fragmentation distributions and to the coefficient functions for
inclusive deep-inelastic scattering and semi-inclusive $e^+e^-$ annihilation.
We present an overview of the methods which allow, in many cases, to derive the
coefficients of the highest three logarithms at all orders in the strong 
coupling from next-to-next-to-leading order results in massless perturbative
QCD. Some representative analytical and numerical results are shown, and the 
present limitations of these resummations are discussed.}

\FullConference{``Loops and Legs in Quantum Field Theory'', 
11th DESY Workshop on Elementary Particle Physics \\
April 15-20, 2012, Wernigerode, Germany}

\newcommand{\beq}{\begin{equation}}
\newcommand{\eeq}{\end{equation}}
\newcommand{\bea}{\begin{eqnarray}}
\newcommand{\eea}{\end{eqnarray}}
\newcommand{\nn}{\nonumber}
\newcommand{\ra}{\rightarrow}

\newcommand{\hspn}{{\hspace{-3mm}}}
\newcommand{\hspp}{{\hspace{3mm}}}

\newcommand{\MSb}{$\overline{\mbox{MS}}$}
\newcommand{\as}{\alpha_{\,\sf s}}
\newcommand{\ass}{\alpha_{\,\sf s}^{\,2}}
\newcommand{\asth}{\alpha_{\,\sf s}^{\,3}}
\newcommand{\asf}{\alpha_{\,\sf s}^{\,4}}
\newcommand{\ar}{a_{\sf s}}
\newcommand{\ars}{a_{\sf s}^{\,2}}
\newcommand{\art}{a_{\sf s}^{\,3}}

\newcommand{\atil}{\tilde{a}_{\rm s}}
\newcommand{\Ntil}{\widetilde{\!N}}
\newcommand{\GE}{\gamma_{\rm e}}

\newcommand{\ep}{\epsilon}

\def\z#1{{\zeta_{#1}^{}}}

\def\zt2{{\zeta_{2}^{\,3}}}
\def\zf2{{\zeta_{2}^{\,4}}}

\def\ca{{C_{\!A}}}
\def\cas{{C^{\, 2}_{\!A}}}
\def\cath{{C^{\, 3}_{\!A}}}
\def\cafo{{C^{\, 4}_{\!A}}}
\def\cf{{C_F}}
\def\nf{{n^{}_{\! f}}}
\def\nfs{{n^{2}_{\! f}}}

\def\cfs{{C^{\, 2}_F}}

\def\caf{{C_{AF}}}
\def\cafs{{C_{AF}^{\: 2}}}

\def\b#1{{{\beta}_{#1}^{}}}
\def\bb#1#2{{{\beta}_{#1}^{\,#2}}}
\def\B#1{{{\cal B}_{\:\!#1}}}

\def\frct#1#2{\mbox{\Large{$\frac{#1}{#2}\:\!$}}}
\def\frkt#1#2{\mbox{\large{$\frac{#1}{#2}\:\!$}}}

\def\x1{{(1\! -\! x)}}

\begin{document}

\section{Introduction: splitting and coefficient functions and their endpoint
 behaviour}

\noindent
(Semi-)$\,$inclusive lepton-hadron processes, see Refs.~\cite{FP82,PDG12}, 
provide benchmark observables in high-energy lepton-nucleon, electron-positron 
and proton-(anti-)$\,$proton collisions. 
We mainly report on the first two cases here, and specifically consider 
structure functions in inclusive deep-inelastic scattering (DIS) and 
fragmentation functions in single-hadron inclusive (semi-inclusive) 
electron-positron annihilation (SIA), 
$\,e^+e^- \ra\, \gamma^{\,\ast},\:Z,\:H(q) \,\ra\, h(p) + X$, 
in massless perturbative~QCD. 

\vspace*{1mm}
Disregarding contributions suppressed by powers of $Q$, these one-scale 
quantities are given~by
\beq
\label{Fa-pQCD}
 F_a^{\,h}(x,Q^2) \; = \;
 \big[ \, C_{a,i\,}(\as(\mu^2),\mu^2\!/Q^2) \otimes f_{i}^{\,h} (\mu^2)
 \big] \!(x) 
\eeq
in terms of their {\it coefficient functions} $C_{a,i}$ and the parton or
fragmentation distributions $f_{i}^{\,h}$ of the hadron $h$. 
Here $Q^2$ is the physical hard scale, $Q^2= \sigma\:\! q^2$ with $\sigma= -1$ 
for DIS and $\sigma = 1$ for SIA, where $q$ is the momentum of the exchanged 
gauge or Higgs boson, $x$ is the corresponding scaling variable, $x = 
[(2p\cdot q)/Q^2]^\sigma$, and $\otimes$ abbreviates the Mellin convolution. 
The dependence of $f_{i}^{\,h}$ on the renormalization and factorization scale 
$\mu$ is given by the renormalization-group equations
\beq
\label{Evol}
  \frct{d}{d \ln \mu^2} \: f_i^{}(x,\mu^2) \; =\;
  [ \, P^{\,S,T}_{ik,\,ki}(\as(\mu^2)) \otimes f_k^{}(\mu^2) \,](x)
  \:\: ,
\eeq
where $P^{\,S,T}$ are the `spacelike' ($\sigma=-1$) and `timelike' ($\sigma=1$)
{\it splitting functions} and $x$ now represents momentum fractions.
Appropriate summations over $i$ in Eq.~(\ref{Fa-pQCD}) and $k$ in  
Eq.~(\ref{Evol}) are understood.
Choosing $\mu^2 = Q^2$ without loss of information, the $\as$-expansions of 
$C_{a}$ and $P$ read
\beq
\label{PCexp}
  C_{a}(x,\as) \:=\: 
  { \mbox{\small $\displaystyle \sum_{\,n\,=\,0}^{}$ }}\! 
  \ar^{\,n+n_a}\, c_{a}^{(n)}(x) \;\; , \quad
  P(x,\as) \:=\: 
  { \mbox{\small $\displaystyle \sum_{\,n\,=\,0}^{}$ }}\! 
  \ar^{\,n+1} P^{\,(n)}(x) \quad \mbox{ with } \quad
  0 < x < 1 \;\; .
\eeq

\vspace*{-1.5mm}
\noindent
We normalize the strong coupling as $\ar \,=\, \as(Q^2)/(4\pi)$.
The contributions up to $c_{a}^{(\ell)}$ and $P^{\,(\ell)\!}$ form the 
N$^\ell$LO (leading order, next-to-leading order, \dots) `fixed-order' 
approximation for $F_a$. 

\vspace*{1mm}
The initial-state splitting functions $P^{\,S}$ and the coefficient functions 
for the DIS structure functions $F_{L,2,3,\phi\,}$, where $F_\phi$ denotes the 
Higgs-exchange structure function in the heavy top-quark limit, are known to 
order $\asth$ from the diagram calculations in 
Refs.~\cite{MVV34,MVV5,MVV6,MVV10,SMVV}. 
The corresponding results for the final-state quantities $P^{\,T}$ have been 
derived only by indirect means so far \cite{P2Tqq,P2Tgg,AMV}; in fact, the 
latter results still include an uncertainty which needs to be addressed by 
future calculations (but is not relevant in the present context). 
The coefficient functions for the fragmentation functions $F_{L,T,A,\phi}$, 
cf.~Ref.~\cite{PDG12}, are completely known only at order $\ass$ 
\cite{RvN1,RvN2,MMoch06,AMV}.

\vspace*{1mm}
Generically, the $\as$-coefficients in Eq.~(\ref{PCexp}) can be expanded around
$x=1$ and $x=0$ in the~form
\beq
\label{DLogs}
  \big\{ c_a^{\,(n)\!},\, P^{\,(n)} \big\}
  \;=\; \mbox{\small $\displaystyle\sum_{r\, = -1}^{\infty}$ }\! X^r 
  \left( A_{r,0} \ln^{\,2n-n_0^{}} X 
       \,+\, A_{r,1} \ln^{\,2n-n_0^{}-1} X 
       \,+\: \ldots\, \right) 
\eeq 
 
\vspace*{-1.5mm}
\noindent
with $X = 1\!-\!x\,$ or $\,X = x$, where $X \!\ll\! 1$ respectively represents 
the threshold (large-$x$) and high-energy (small-$x$) limits. 
The important exceptions to this {\it double-logarithmic enhancement} (i.e., 
two additional powers of $\ln X$ occur per order in $\as$) are the diagonal 
splitting functions $P_{\rm qq}^{\,S,T}$ and $P_{\rm gg}^{\,S,T\!}$, which show
no large-$x$ enhancement at $r=-1$ and $r=0$ in the standard \MSb\ scheme 
\cite{Pqqthr1,DMS05}, 
and the flavour-singlet splitting functions $P^{\,S}$ and DIS coefficient 
functions, which are only single-log (`BFKL') enhanced in the \mbox{$r=-1$} 
terms dominating the small-$x$ behaviour \cite{BFKL1,BFKL2}. 
The dominant $r=-1$ large-$x$ contributions to the coefficient functions can be
resummed in the framework of the soft-gluon exponentiation, see, e.g., 
Refs.~\cite{SGlue1,SGlue2} and for the present status 
Refs.~\cite{sgDIS,BRsia,MV4}.

\pagebreak

\begin{figure}[thb]
\vspace*{-1mm}
\centerline{\hspace*{-2mm}\epsfig{file=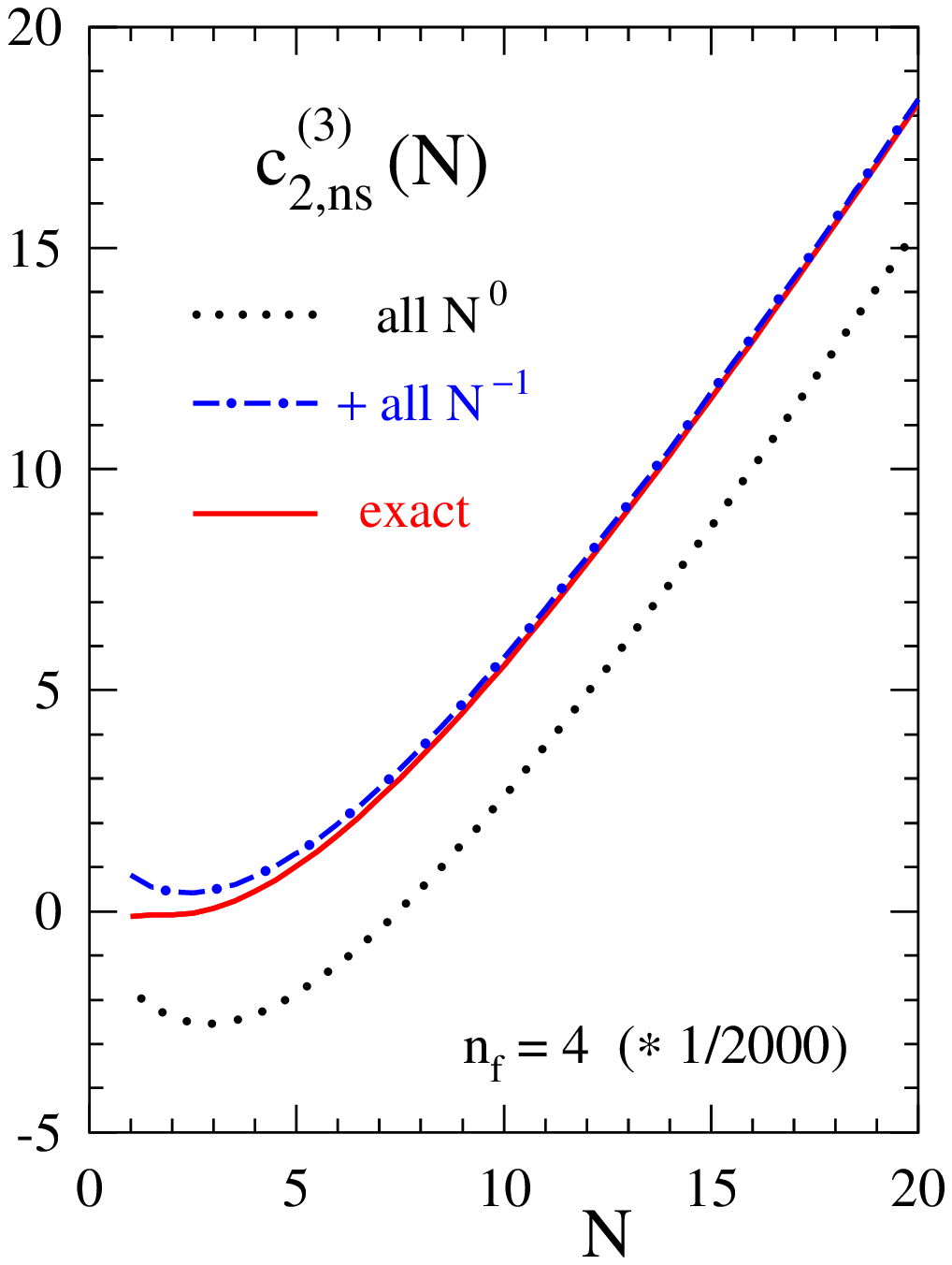,width=6cm,angle=0}%
\hspace*{-2mm}\epsfig{file=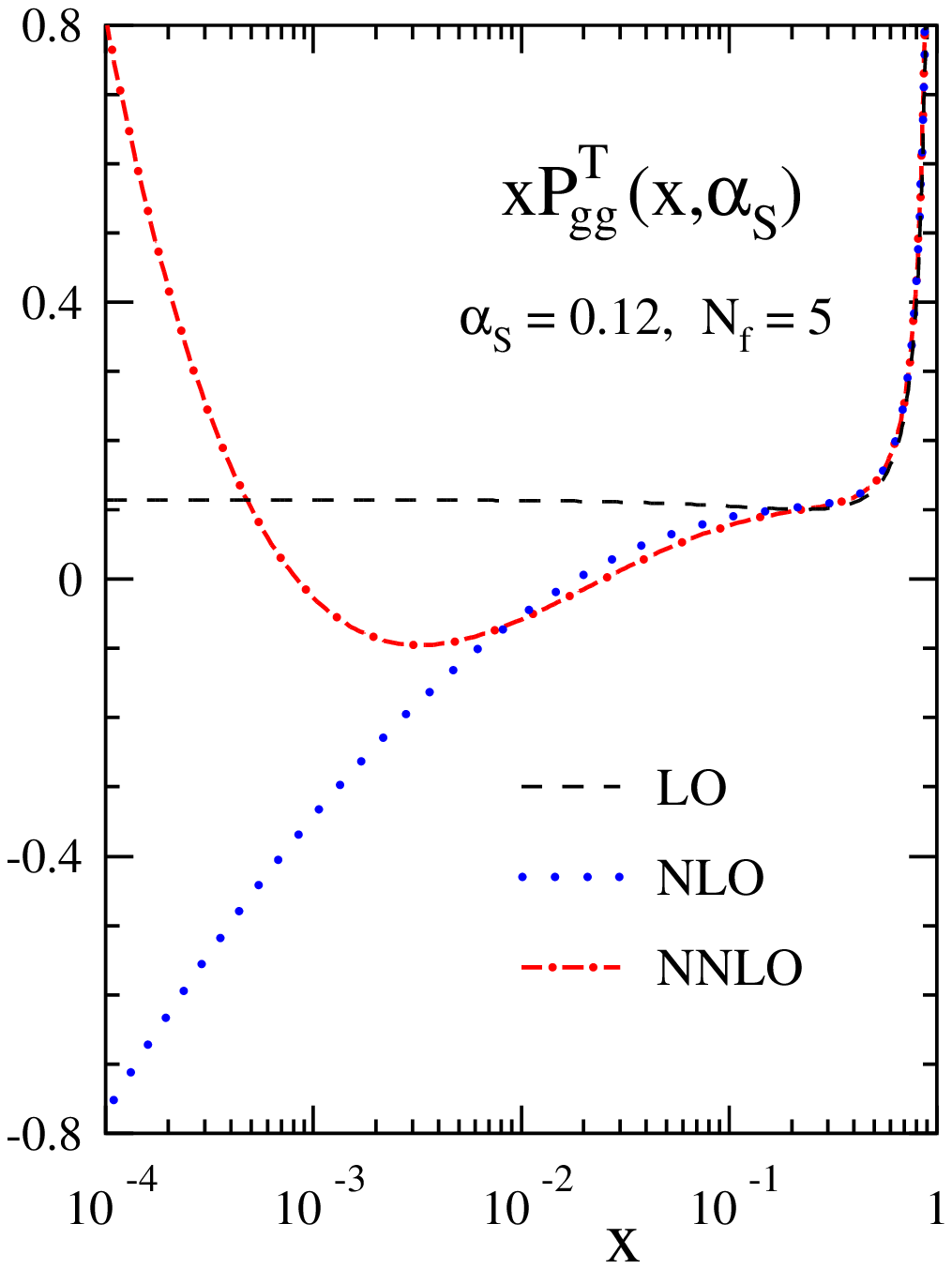,width=5.73cm,angle=0}}
\vspace{-2.5mm}
\caption{ \label{fig1}
 Left: the third-order coefficient function for $F_{2,\rm ns}$ \cite{MVV6} 
 compared to large-$N$ approximations by only the soft$\,$+$\,$virtual 
$N^{\,0\!}$ terms and by those plus the ($\,r=0\,$) $N^{\,-1\!}$ contributions. 
 Right: the LO, NLO and NNLO \cite{P2Tgg} approximations to the timelike 
 splitting function $P_{\rm gg}^{\,T}$ for five flavours at a typical scale 
 $Q^{\,2} \simeq M_{Z\!}^{\:2\!}$.}
\vspace*{-2mm}
\end{figure}

In the remainder of this contribution we briefly summarize results derived
for all other cases in 
Refs.~\cite{MV3,MV5,SMVVs5,LL2010,AV2010,ASV,AV2011,Rcor11,KVY,ALPV,KV1}
with some emphasis on the phenomenologically directly relevant cases of $r=0$ 
at large $x$ in DIS and $r=-1$ at small $x$ in~SIA. 
This relevance is illustrated in Fig.~\ref{fig1}: the left panel shows, after 
transformation to Mellin-$N$ space, that the soft and virtual contributions do 
not sufficiently dominate the N$^3$LO non-singlet quark coefficient function 
for $F_2$ at `large' $N$;
the right panel illustrates the dramatic small-$x$ instability of the known
fixed-order approximations for~$P_{\rm gg}^{\,T}$.

\section{Fourth- and all-order ln$^{\,\rm k}$(1-x) predictions from physical 
kernels}

\noindent
It can be useful to eliminate the parton or fragmentation distributions, and
the associated choice of a factorization scheme and scale, from the 
description of the dependence of observables on the physical scale $Q^2$. 
This leads to {\it physical evolution kernels} $K_a$, which generically can be
written~as
\beq
\label{Ka-pQCD}
  \frct{d F_a^{\,h}}{d \ln Q^2} \;=\;
  \frct{d}{d \ln Q^2} \, \big( C_a \otimes f^{\,h\,} \big) \; = \;
  \big( \beta(\ar)\, \frct{d\:\! C_a}{d \ar}\, + \, C_a \otimes P \big)\, 
  \otimes C_a^{\,-1} \otimes F_a^{\,h} 
  \; \equiv \; K_a \otimes F_a^{\,h} \:\: ,
\eeq
where $\beta(\ar) \,=\, - \beta_0\, \ars - \beta_1 \art - \dots\,$, with
$\beta_0 = \frac{11}{3}\: \ca - \frac{2}{3}\: \nf\,$ in our normalization, is 
the (four-dimensional) beta function of QCD. $\ca = 3$ and $\cf = \frac{4}{3}$
are the usual SU($n_{\it colours}\!=\!3$) group factors, and $\nf$ denotes 
the number of effectively massless flavours. 

\vspace*{1mm}
For {\bf flavour non-singlet cases} such as, e.g., the structure function 
$F_3$, (\ref{Ka-pQCD}) is a scalar equation. The first term then represents a 
logarithmic derivative in $N$-space, and the second is reduced to a
combination of quark-(anti-)$\,$quark splitting functions. The resummations 
discussed here were initiated by observing that the $K_{a,\rm ns}$ in DIS are 
{\it single-log enhanced} not only for the $\x1^{\,-1}$ terms, as guaranteed by 
the soft-gluon exponentiation, cf.~Ref.~\cite{NV3}, but at all powers in~$\x1$,
\beq
\label{Ka-Logs}
  K_a\big|_{\ar^{\,n+1}}
  \;=\; \mbox{\small $\displaystyle\sum_{r\, = -1}^{\infty}$ }\! \x1^r
  \big( K_a^{\,(r,0)} \ln^{\,n} \x1 \,+\, K_a^{\,(r,1)} \ln^{\,n-1} \x1 
  \,+\: \ldots\, \big) \:\: .
\eeq
If this behaviour holds also beyond the orders covered by 
Refs.~\cite{MVV5,MVV6,MVV10}, as suggested by the all-order leading 
large-$\nf$ result of Refs.~\cite{F2Lnf}, it implies a {\it resummation of the 
double-logarithmic contributions} to the coefficient functions $C_{a,\rm ns}$ 
due to the non-enhancement of $P_{\rm qq}$ mentioned above.

\vspace*{0.5mm}
A closer look, see also Ref.~\cite{Rcor09} for an introductory overview, 
reveals that the third-order coefficient functions are sufficient to fix the 
coefficients of the {\it highest three logarithms} in Eq.~(\ref{DLogs}) for 
(the non-singlet components of) $C_{a,\rm q}$.
The resulting all-order $N^{\,-1}$ [$r = 0\,$ in Eq.~(\ref{DLogs})] predictions
have been presented in Ref.~\cite{MV3} for $a=L$ [with $n_L^{} = 1$ and 
$n_0^{} = 0$ in Eqs.~(\ref{PCexp}) and (\ref{DLogs})] and in section 5 of 
Ref.~\cite{MV5} for $a = 1,2,3$ [where $n_a = 0$ and $n_0^{} = 1$]. 
In the latter cases, only one parameter, denoted by $\xi_{\rm DIS_4}^{}$, is
missing for the fourth logarithms. 
Section 5 of Ref.~\cite{MV5} also includes the highest three logarithms at 
order $\asf$ for all these cases at all powers in $1\!-\!x$, written down in a 
closed form using suitably modified harmonic polylogarithms \cite{HPLs} up to 
weight~3. 

\vspace*{0.5mm}
Completely analogous results for the fragmentation functions $F_{L,T,I,A}$ in 
SIA can be found in section 6 of Ref.~\cite{MV5}. As mentioned above, the
third-order coefficient functions are not fully known for these quantities. 
However, it was possible to derive the highest three \mbox{large-$x$} 
logarithms to all orders in $1\!-\!x$ using the (generally non-trivial) 
analytic-continuation (${\cal AC}$) or Drell-Yan-Levy relation
between inclusive DIS and SIA, cf.~also Ref.~\cite{BRvN00} and references
therein; the results are given in section 3 of Ref.~\cite{MV5}.
Finally the same approach can be applied to the quark-antiquark annihilation
contribution to the total cross section for 
{\it Drell-Yan lepton-pair production},
$pp/p\bar{p} \,\ra\, l^+l^-+X$. In this case the fixed-order information is
limited to order $\ass$ \cite{DYnnlo}, hence only the coefficients of the 
{\it highest two logarithms} in Eq.~(\ref{DLogs}) are completely
determined by the single-logarithmic enhancement of the corresponding physical 
kernel. The resulting all-order $N^{\,-1}$ and third- and fourth-order all-$r$
predictions can also be found in section 6 of Ref.~\cite{MV5}. 

\vspace*{0.5mm}
The single-log enhancement (\ref{Ka-Logs}) also holds for the $2\times 2$ 
matrices of {\bf flavour-singlet kernels} for combinations such as $F_{2\phi} 
= (F_2,\,F_\phi)$ and $F_{2L} = (F_2,\,F_L)$ in DIS and their  counterparts in 
SIA. In~these cases both the coefficient and splitting functions 
contribute double-log enhanced terms on the r.h.s.~of Eq.~(\ref{Ka-pQCD}), 
hence Eq.~(\ref{Ka-Logs}) does not imply an all-order resummation. 
It is, however, possible to use this equation for $F_{2\phi}$ at $n=3$ to 
derive the highest three large-$x$ logarithms of the (otherwise unknown) 
N$^3$LO {\it singlet splitting functions} $P^{\,(3)S}_{ik}$ from the N$^3$LO 
coefficient functions computed in Refs.~\cite{MVV6,SMVV}. 
This calculation has been carried out in Ref.~\cite{SMVVs5} to all orders in 
$r$. A~surprising, at the time, outcome was that the coefficient of all four 
leading logarithms, $\ln^{\,6} \x1$ for $P^{\,(3)}_{i\neq k}(x)$ and 
$\ln^{\,5} \x1$ for $P^{\,(3)}_{i=k}(x)$, were found to vanish to all orders 
in $1\!-\!x$. 

\vspace*{0.5mm}
With the highest three logarithms determined for $P^{\,(3)S\!}_{ik}$, it became 
possible to employ Eq.~(\ref{Ka-Logs}) for $F_{2L}$ to derive corresponding 
results for the [four-loop, due to $n_a = 1$ in Eq.~(\ref{PCexp})] N$^3$LO 
singlet {\it coefficient functions for} $F_L$. This has been done in Ref.\
\cite{LL2010}, but only for the leading-$r$ contributions, with new results for 
$c_{L,\rm g}^{\,(3)}$ at $r=1$. The all-$r$ generalization, which would also
yield results beyond those of Refs.~\cite{MV3,MV5} also for the quark 
coefficient function, is not available in the literature~yet. 

\vspace*{0.5mm}
No new large-$x$ results have been derived (only) from physical evolution 
kernels after Ref.~\cite{LL2010}. The limitations of this approach 
-- the need to conjecture the all-$n$ validity of Eq.~(\ref{Ka-Logs}) and lack 
of all-order predictions for the singlet splitting and coefficient functions --
have been overcome since then by starting from the unfactorized structure and 
fragmentation functions as discussed below. However, the physical kernels still
represent the easiest route to all-$r$ results at order $\asf$, and they 
can provide invaluable hints for the functional forms of the all-order 
large-$x$ coefficient functions.

\section{Double logarithmic endpoint resummations via unfactorized quantities}

\noindent
The splitting functions and the coefficient functions for the (combinations of)
DIS and SIA observables discussed above originate in the {\it unfactorized
expressions} in $D = 4 - 2\bar{\epsilon}$ dimensions,
\beq
\label{Fhat}
  \widehat{F}_a \;=\; \widetilde{C}_a \otimes Z \quad \mbox{ with } \quad
  P \;=\; \beta_D^{}(\ar) \,\frct{dZ}{d\ar}\, \otimes Z^{-1} \; .
\eeq
Here $\widetilde{C}_a$, the $D$-dimensional coefficient functions, are given 
by Taylor series in $\ep$ 
(which differs from $\bar{\ep}$ by some `artefacts' of dimensional 
regularization)
with the $\ep^{\,0}$ coefficient leading to the `physical' (\MSb-scheme) 
coefficient function in Eqs.~(\ref{Fa-pQCD}) and (\ref{PCexp}).  
$\beta_D^{}(\ar)$ is the $D$-dimensional beta function, 
$\beta_D^{}(\ar) \,=\, - \ep \ar + \beta(\ar)$ with $\beta(\ar)$ defined below
Eq.~(\ref{Ka-pQCD}). For non-singlet cases all quantities in Eq.\ 
(\ref{Fhat}) are scalars, for flavour-singlet combinations $\widehat{F}_a$,
$\widetilde{C}_a$, $P$ and $Z$ are $2\times 2$ matrices [unlike in 
Eq.~(\ref{Fa-pQCD}), the quark and gluon contributions to $F_a$ are considered
separately in Eq.~(\ref{Fhat})]. $Z$ consists of $1/\ep$ {\it poles up to} 
$\ep^{\,-n}$ at order $\ar^{\,n}$, and its $\as^{\,n} \ep^{\,-n+\ell}$ 
coefficients include endpoint logarithms up to $\ln^{\,n+\ell-1} X$ with 
$X = 1\!-\!x$ and $X = x$. Hence we have, generically,
\beq
\label{FhatLogs}
  \widehat{F}_{a} \big|_{\ar^{\,n}\:\!\ep^{\,-n+\ell}} \;=\;
  \mbox{\small $\displaystyle\sum_{r\, = -1}^{\infty}$ }\! X^r \,\big( \,
  {\cal F}_{\!a,n,\ell,r}^{(0)} \ln^{\:\!n+\ell-1} X \:+\:
  {\cal F}_{\!a,n,\ell,r}^{(1)} \ln^{\:\!n+\ell-2} X \:+\: \dots \big) \;\; .
\eeq

The second equation in (\ref{Fhat}) can be iteratively inverted to yield $Z$ in 
terms of the expansion coefficient of $P(x,\as)$ and $\beta(\ar)$. The
resulting dependence on $P^{\,(n)}$ and $\beta_n$ can be summarized~as
\beq
\label{ZofPn}
  \ar^{\,n}\ep^{\,-n}:\; P^{\,(0)\!},\: \b0 \; , \quad
  \ar^{\,n}\ep^{\,-n+1}:\; +\, P^{\,(1)\!},\: \b1 \; , \quad
  \ar^{\,n}\ep^{\,-n+2}:\; +\, P^{\,(2)\!},\: \b2 \; , \quad
  \dots \; , \quad
  \ar^{\,n}\ep^{\,-1}:\; P^{\,(n-1)} \; .
\eeq
Explicit all-order expressions for $r=0$, recall Eq.~(\ref{DLogs}), in the 
large-$x$ case have been presented in section 2 of Ref.~\cite{ASV}. 
Beyond this accuracy and for the small-$x$ case, $Z(\as,\ep)$ has been 
evaluated iteratively to a high but finite order in $\as$. 
Eq.~(\ref{ZofPn}) implies that a fixed-order knowledge at N$^m$LO, recall 
Eq.~(\ref{PCexp}), determine the first $m\!+\!1$ non-vanishing coefficients in 
the $\ep$ expansion of $\widehat{F}_{a}$ at all orders in~$\ar$.
If this determination can be extended, at a certain logarithmic accuracy, to 
all orders in~$\ep$, then the {\it all-order mass factorization} can be 
performed at this accuracy, yielding a resummation of the splitting and 
coefficient functions.
In practice this is done order by order in $\as$; we have employed recent 
developments in {\sc Form} \cite{FORM} to reach as high an order as possible.

\vspace*{0.5mm}
One way to obtain all-$n$ and all-$\ep$ expressions for coefficients 
${\cal F}^{(k)}$ in Eq.~(\ref{FhatLogs}) is by expressing the coefficients 
$\widehat{F}_{a}^{(n)}$ in terms of known lower-order results, i.e., by 
finding an all-order {\bf iteration of the unfactorized expressions} 
$\widehat{F}_{a}^{(n)}$ at a given logarithmic accuracy. This approach has 
been used for the $r=0$ {\it large-$x$ structure functions} in 
Refs.~\cite{AV2010,ASV} at $k=0$ and $k=1$. For example, the unfactorized 
gluon contributions to the structure function $F_2$ can be expressed~as
\beq
\label{F2Git}
 \widehat{F}_{2,\rm g}^{\,(n)} \;=\;
 { 1 \over n } \: \widehat{F}_{2,\rm g}^{\,(1)}
 \bigg\{ \,\sum_{i=0}^{n-1}\, f_{2,\rm g}^{}(n,i,\ep)\,
  \widehat{F}_{\phi,\rm g}^{\,(i)} \, \widehat{F}_{2,\rm q}^{\,(n-i-1)}
  \;-\; {\beta_0 \over \ep}\: \sum_{i=0}^{n-2}\, g_{2,\rm g}^{}(n,i)\,
  \widehat{F}_{\phi,\rm g}^{\,(i)} \, \widehat{F}_{2,\rm q}^{\,(n-i-2)} 
  \bigg\}
\eeq
with
\beq
\label{F2Gfg}
  f_{2,\rm g}^{}(n,i,\ep) \;=\;
  \bigg( \begin{array}{c} n\!-\!1 \\[-0.5mm] i \end{array} \bigg)^{-1}
  \,\left[ \, 1 \:+\: \ep f_{2,\rm g}^{(1)}(n,i) \right] \;\; , \quad
  g_{2,\rm g}^{}(n,i) \;=\; 
  \bigg( \begin{array}{c} n \\[-0.5mm] i+\!1  \end{array} \bigg)^{-1}
\eeq
at next-to-leading logarithmic (NLL) accuracy. The $r\!=\!-1$ `diagonal' 
quantities $\widehat{F}_{2,\rm q}^{\,(k)}$, $\widehat{F}_{\phi,\rm g}^{\,(k)}$
required in Eq.~(\ref{F2Git}) are known to a high accuracy from the soft-gluon 
exponentiation, see Ref.~\cite{ASV}. 
 
Especially the non-LL parts of this result (the 
function $f_{2,\rm g}^{(1)}(n,i)$ can be found in section 3 of Ref.~\cite{ASV})
have rather been `engineered' and verified -- using results in Section 2 the 
ensuing system of linear equations is overconstrained by two equations per 
order in~$\as$ -- than derived from first principles. 
Since a second approach turned out to be clearer and more convenient, we have 
not attempted to generalize Eq.~(\ref{F2Git}) to the third (NNL) logarithms. 
Still, the iteration of unfactorized observables might offer access to 
resummations of quantities outside the scope of 
Refs.~\cite{ASV,AV2011,Rcor11,KVY,ALPV,KV1}.

\vspace{0.5mm}
That second approach is to {\bf decompose the $n^{\,th}$-order unfactorized 
expressions} at $x < 1$, without any reference to lower-order counterparts, 
in $n$ terms of $D$-dimensional exponentials, e.g.,
\beq
\label{FThatDec}
  \big( A_{n,k} \,\ep^{\,-2n+1} \,+\, B_{n,k} \,\ep^{\,-2n+2}
  \,+\, C_{n,k} \,\ep^{\,-2n+3} \,+\, \ldots \big) 
  \, X^{r-(\eta_0^{} + k\,\eta_1^{}) \ep}
  \;\; , \quad k = 1,\, \dots,\, n \;\; .
\eeq
For the {\it large-$x$ logarithms}, $X = 1\!-\!x$, this structure arises at 
all $r$ in inclusive DIS and SIA from the phase-space integrations for the 
undetected final-state particles and the loop integrals of the virtual 
corrections \cite{MMvN,RvN2} with $\eta_0^{} = 0$ and $\eta_1^{} = 1$. 
The decomposition (\ref{FThatDec}) is related, but not identical to that into 
contributions with $1,\, \dots,\, n$ undetected partons in the final state. 
In the soft-gluon limit, $r=-1$, it has been employed before to 
`reverse-engineer' the $\asth \ep^{\,-n}$ pole terms of the $\gamma^{\ast}qq$ 
and $Hgg$ form factors (and the $\nf \asth \ep^{\,0}$ contributions in the 
former case) from the calculations for Refs.~\cite{MVV6,SMVV} in 
Refs.~\cite{MVV89} and to extend the soft-gluon resummation to 
N$^3$LL accuracy for SIA, Drell-Yan lepton-pair production and the total cross 
section for Higgs production via gluon-gluon fusion \cite{MV3,MV1}. 
It may be worthwhile to note that the results of Refs.~\cite{MVV89} were
confirmed (and substantially extended, to order $\asth \ep^{\,2}$) by 
direct diagram calculations in Refs.~\cite{FF3a,FF3b}.

\vspace{0.5mm}
The situation is far more complicated for the {\it small-$x$ logarithms}, 
$X = x$.
Here we focus on contributions that do not vanish for $x \ra 0$, i.e., $r=0$ 
and $r=-1$. In the former case Eq.~(\ref{FThatDec}) is found to hold with 
$\eta_0^{} = 0$ and $\eta_1^{} = -1$ in DIS and $\eta_0^{} = 1$ and 
$\eta_1^{} = 1$ in SIA, as one might have `naively' expected from 
Refs.~\cite{MMvN,RvN2}, but only for `even-$N$' quantities such as 
$F_{L,1,2,\phi}$ in DIS and their SIA counterparts. 
For $g_1^{}$ in polarized DIS and $F_3$, where the operator-product expansion
provides the odd Mellin moments, and the corresponding asymmetric fragmentation
function $F_A$ already the colour structure of the leading logarithms 
\cite{MVV10,BVns,BVpol} excludes such a decomposition.
On the other hand, Eq.~(\ref{FThatDec}) is applicable all orders at $r=-1$ in 
SIA, but with $\eta_0^{} = 0$ and $\eta_1^{} = 2$, another structure that has 
been discovered and verified but still lacks a proper explanation.

\vspace{0.5mm}
After expanding in $\ep$, the $\ep^{\,-2n+1}, \:\dots\,,\: \ep^{\,-n-1}$ terms
in Eq.~(\ref{FThatDec}) have to cancel in the sum~(\ref{Fhat}).
This implies $\,n\!-\!1$ relations between the LL coefficients $A_{n,k}$ which 
lead to the constants ${\cal F}_{n,\ell}^{(0)}$ in Eq.~(\ref{FhatLogs}), 
$\,n\!-\!2$ relations between the NLL coefficients $B_{n,k}$ determining 
${\cal F}_{n,\ell}^{(1)}$ etc.
As discussed above, a N$^m$LO calculation fixes the (non-vanishing)
coefficients of $\ep^{\,-n}, \:\dots\,,\: \ep^{\,-n+m}$ at all orders $n$,
adding $m+1$ more relations between the coefficients in Eq.~(\ref{FThatDec}).
Consequently the highest $m+1$ double logarithms, i.e., {\it the} N$^m$LL
{\it approximation, can be determined from the} N$^m$LO {\it results} wherever
an equation like (\ref{FThatDec}) (the expression for $F_L$, e.g., is slightly
different) is applicable.

\vspace{0.5mm}
Therefore the present fixed-order results allow the determination, to all 
orders in $\as$, of the highest four \mbox{$r=0$} large-$x$ logarithms for the 
coefficient functions in non-singlet DIS (where the accuracy effectively is 
N$^3$LO due to the non-enhancement of the splitting functions) and SIA (due to 
its close relation to the DIS case, see Ref.~\cite{MV5}), and of the 
corresponding highest three \mbox{large-$x$} and small-$x$ logarithms in the 
flavour singlet splitting and coefficient functions, including the 
phenomenologically most important $x^{\,-1}$ terms in SIA illustrated above.

\section{Selected recent results on large-$x$ and small-$x$ double logarithms}
  
\vspace*{-1mm}
\noindent
We now show a small subset of the results derived by the methods described in 
Section 3. In the non-singlet large-$x$ cases the physical-kernel results
have been verified, and the previously missing parameter has been 
determined as $\xi_{\rm DIS_4}^{} = \xi_{\rm SIA_4}^{} = \frkt{100}{3}$,
a result that has been obtained independently in Ref.~\cite{Grunberg:2009vs}.
Except for the numerical impact of the thus known four $r=0$ logs on the
fourth-order coefficient function $C_{2,\rm ns}^{\,(4)}$, which will be 
illustrated in Fig.~2 below, we will focus on the large-$x$ singlet and 
small-$x$ cases, where progress has been made since the Wernigerode~workshop.

\vspace{0.5mm}
The highest three $r=0$ {\it large-$x$ logarithms} in the off-diagonal splitting
functions $P^{\,S}$ and coefficient functions for $F_{L,2,\phi}$ in DIS have 
now been cast in closed all-order forms that supersede the tables in the 
appendix of Ref.~\cite{ASV}. Also the NNLL contributions can be expressed 
in $N$-space in terms of the apparently new {\it Bernoulli functions} $\B{n}(x)$ introduced 
and discussed in Refs.~\cite{AV2010,ASV},
\beq
\label{Bkdef}
  \B{k}(x) \;=\;
  \sum_{n\,=\,0}^\infty 
  \;\frac{B_n}{n!(n+k)!}\; x^{\,n}
 \quad \mbox{and} \quad
  \B{-k}(x) \;=\;
  \sum_{n\,=\,k}^\infty \;\frac{B_n}{n!(n-k)!}\; x^{\,n} \; ,
\eeq
where $B_n$ are the Bernoulli numbers in the normalization of 
Ref.~\cite{AbrSteg}.
As an example, we show the NNLL approximation to the spacelike gluon-quark 
splitting function which can be written as 
\bea
\label{Pqgxto1}
 \lefteqn{ N P_{\rm qg}^{\,S}(N,\as) \;=\; 2\:\!\ar\,\nf\,\B0(\atil) }
 \nn \\ & & \mbox{}
 +\,\ar^{\,2} \ln\, \Ntil\: \nf
  \Big\{ \left( 6\, \cf-\b0 \right) 
        \left( \B1(\atil) + 2\,\atil^{\,-1} \B{-1}(\atil) \right)
  \,+\: \b0\, \atil^{\,-1} \B{-2}(\atil) \Big\}
 \\[-1mm] & & \mbox{}
 +\, \frct{\ar^2 \,\nf}{48\*\,\caf}\*
        \Big\{
                  \bb02\* \Big[
                          2 \* \atil \* \B2(\atil)
                        - 12 \* \B1(\atil)
                        + 12 \* \B0(\atil)
                        - 6 \* \B{-1}(\atil)
                        - 12 \* \atil^{\,-1} \* \B{-2}(\atil)
\nn \\[-1mm] & & \mbox{\hspp}
                        - 4\, \* \atil^{\,-1} \* \B{-3}(\atil)
                        + 3\, \* \atil^{\,-1}\* \B{-4}(\atil)
                  \Big]
                \, - \,36 \*\, \b0\* \cf \* \Big[
                          \atil \* \B2(\atil)
                        - \,3 \* \B1(\atil)
                        + 4 \* \B0(\atil)
                        -  \B{-1}(\atil)
\nn \\[-0.5mm] & & \mbox{\hspp}
                        + \atil^{\,-1} \* \big(
                          2 \* \B{-1}(\atil)
                        + \B{-2}(\atil)
                        - 2 \* \B{-3}(\atil) \big)
                  \Big]
                \, + \, 108\* \, \cfs\* \Big[
                          2\, \* \atil \* \B2(\atil)
                        - \,4 \* \B1(\atil)
                        + 5 \* \B0(\atil)
\nn \\[-0.5mm] & & \mbox{\hspp}
                        + \atil^{\,-1} \* \big(
                          2 \* \B{-1}(\atil)
                        + 4 \* \B{-2}(\atil) \big)
                  \Big]
                \, + \, 80\* \,\caf \* \b0 \* \Big[
                          \atil \* \B2(\atil)
                        - \, 4 \* \B1(\atil)
                        + \, 4 \* \B0(\atil)
                        +  \B{-1}(\atil)
                  \Big]
\nn \\[-0.5mm] & & \mbox{\hspp}
                - 32\* \,\caf\* \,\cf \* \Big[
                          (19-3 \* \z2) \* \atil \* \B2(\atil)
                        - 34 \* \B1(\atil)
                        + (13+6 \* \z2) \* \B0(\atil)
                        - (2-3 \* \z2) \* \B{-1}(\atil)
                  \Big]
\nn \\[-2mm] & & \mbox{\hspp}
                + 32\* \,\cafs\* \Big[
                          (2+3 \* \z2) \* \atil \* \B2(\atil)
                        + (4+12 \* \z2) \* \B1(\atil)
                        + (2-12 \* \z2) \* \B0(\atil)
                        + (2-3 \* \z2) \* \B{-1}(\atil)
                  \Big]
        \!\bigg\}
\nn 
\eea
with $\,\ln\, \Ntil \,=\, \ln N + \GE$, $\atil \,=\, 4\:\!\ar\,\caf\,\ln^2 
\Ntil$ and $\caf = \ca - \cf$, where $\GE$ is the Euler constant.
Corresponding results for $P_{\rm gq}^{\,S}$, their timelike counterparts with,
e.g., $C_F^{\,-1} P_{\rm gq}^{\,T} = n_{\! f}^{\,-1} P_{\rm qg}^{\,S}$ at LL 
accuracy, and the related coefficient functions will be presented 
in Ref.~\cite{ALPV}.
These results explain the vanishing LL coefficients at order $\asf$ as the 
start of an all-order pattern due to $B_{2n+1} = 0$ for $n \geq 1$.

\vspace*{0.5mm}
We now turn to the $r=-1$ {\it small-$x$ resummation in SIA}, where even more 
results were only known via tables of expansion coefficients in 
Ref.~\cite{AV2011}. 
This situation changed dramatically as a consequence of discussions which, in 
all likelihood, would not have occurred without the 2012 Loops \& Legs workshop.
All results of Ref.~\cite{AV2011}, and more, are now known in closed form, 
e.g.,
\bea
\label{PTqq-cl}
  P_{\rm qq}^{\,T}(N) &\, = \,& 
  \frct{4}{3}\,\*\frkt{\cf\*\nf}{\ca}\,\*\ar\*
  \big\{ {\frct{1}{2\*\xi}\* (S-1)\* ({\cal L}+1) + 1} \big\}
\nn \\[-0.5mm] & & \mbox{\hspn\hspn\hspn}
  + \frct{1}{18}\,\*\frkt{\cf\*\nf}{\cath}\,\*\ar\* \bar{N}\* \big\{
    ( - 11\,\*\cas + 6\,\*\ca\*\nf - 20\,\*\cf\*\nf )\,\*
    { \frct{1}{2\*\xi}\*(S-1+2\,\*\xi) }
  + 10\,\*\cas\,\* { \frct{1}{\xi}\*(S-1)\*{\cal L} }
\nn \\[-1.2mm] & & \mbox{\hspn\hspn}
  - ( 51\,\*\cas - 6\,\*\ca\*\nf + 12\,\*\cf\*\nf )\*\, { \frct{1}{2}\*(S-1) }
  + ( 11\,\*\cas + 2\,\*\ca\*\nf - 4\,\*\cf\*\nf )\*\, {  S^{\,-1}\* {\cal L} }
\nn \\[-0.2mm] & & \mbox{\hspn\hspn}
  + ( 5\,\*\cas - 2\,\*\ca\*\nf + 6\,\*\cf\*\nf )\*\,
    { \frct{1}{\xi}\* (S-1)\* {\cal L}^2 }
  + ( 51\,\*\cas - 14\,\*\ca\*\nf + 36\,\*\cf\*\nf )\* {\cal L} \big\} \; , 
\eea
\bea
\label{PTgg-cl}
  P_{\rm gg}^{\,T}(N) &\, = \,&
  \frct{1}{4}\* { \bar{N}\* (S-1) }
  \;-\;  P_{\rm qq}^{\,T}(N) 
  \;-\; \frkt{1}{6\:\!\*\ca}\,\* \ar\*\, ( 11\,\*\cas + 2\,\*\ca\*\nf
  - 4\,\*\cf\*\nf )\*\, { (S^{\,-1}-1) }
\nn \\[0.5mm] & & \mbox{\hspn\hspn\hspn} + \:
  \frkt{1}{576\*\,\cath}\*\:\ar\* \bar{N}\* \big\{ \!\!
   \left( [1193 - 576\,\*\z2]\* \cafo - 140\,\*\cath\*\nf + 4\,\*\cas\*\nfs
   - 56\,\*\cas\*\cf\*\nf - 48\,\*\cfs\*\nfs
\right. \quad \nn \\[-0.5mm] & & \left. \mbox{\hspn\hspn}
   + 16\,\*\ca\*\cf\*\nfs \right)\* { (S-1) }
   \,+\, \left( [830 - 576\*\,\z2]\*\,\cafo + 96\*\,\cath\*\nf
   - 8\*\,\cas\*\nfs - 208\*\,\cas\*\cf\*\nf
\right.  \nn \\[1mm] & & \left. \mbox{\hspn\hspn}
   + 64\*\,\ca\*\cf\*\nfs - 96\*\,\cfs\*\nfs \right)\* { (S^{\,-1}-1) }
   \:+\: ( 11\*\,\cas + 2\*\,\ca\*\nf - 4\*\,\cf\*\nf )^2 \* { (S^{\,-3}-1) }
  \big\} 
\eea
with
$ 
  S \,=\, ( 1 - 4\:\!\xi )^{1/2} \, , \;
  {\cal L} \,=\, \ln \big(\, \frct{1}{2}(1+S) \big) \, , \;
  \xi \,=\, -\:\!8\,\ca\ar \bar{N}^{\,-2} \; \mbox{and} \;
  \bar{N} \,\equiv\, N\!-\!1 \; .
$
The first term in Eq.~(\ref{PTgg-cl}) is the well-known leading-logarithmic 
result of Refs.~\cite{LLxto0}. Beyond this accuracy, the splitting functions
were unknown in the \MSb\ scheme, see Refs.~\cite{ABKK11}, before 
Ref.~\cite{AV2011}. The crucial step towards the closed forms illustrated 
above was the derivation of the first line of Eq.~(\ref{PTqq-cl}) which made 
use of Ref.~\cite{OEIS} as described in appendix A of Ref.~\cite{KVY}. 

\vspace*{0.5mm}
Expressions such as Eq.~(\ref{PTgg-cl}) allow the evaluation of the oscillating
combined NLO$\,$+$\,$NNLL splitting functions down to extremely small $x$,
see Fig.~2, and at $N = 1$ in Mellin space, e.g., 
\bea
\label{PTgg-N1}
  P^{\,T}_{\rm gg}(N=1) &\, = \,& 
  (2\:\!\*\ca\* \ar)^{1/2}
  \,-\, \frct{1}{6\:\!\*\ca}\*
    ( 11\,\*\cas + 2\,\*\ca\*\nf + 12\,\*\cf\*\nf )\,\*\ar
\nn \\[-0.5mm] & & \mbox{\hspn}
  \,+\, \frct{1}{144\,\*\cath}\*
    \left( [ 1193 - 576\,\*\z2 ] \,\*\cafo - 140\,\*\cath\*\nf
      + 4\,\*\cas\*\nfs + 760\,\*\cas\*\cf\*\nf
\right. \nn\\[-1mm] & & \left. \mbox{\hspp\hspp\hspp\hspp}
      - 80\,\*\ca\*\cf\*\nfs + 144\,\*\cfs\*\nfs \right)
    \,\* (2\:\!\*\ca\* \art)^{1/2}
  \;\; + \;\; {\cal O}(\ars) 
\nn \\ &\, \cong \,&
  0.6910\,\as^{1/2} \:-\: 0.9240\, \as \:+\: 0.6490\, \as^{\,3/2}
  \; + \; {\cal O}(\ass) 
  \quad \mbox{for} \quad \nf \,=\, 5 \; . \quad
\eea
The latter results have already been used in an analysis of multiplicities in 
quark and gluon jets \cite{BKK12}. \linebreak See Ref.~\cite{KVY} for more 
results on timelike splitting functions and SIA coefficient functions,
including a first step towards a higher logarithmic accuracy for 
$P^{\,T}_{\rm gg}$ based on Ref.~\cite{DMS05}, see also Ref.~\cite{March06}.

\vspace*{0.5mm}
Similar results have been derived for the {\it small-$x$ logarithms of 
even-$N$ DIS quantities}, e.g., 
\bea
\label{Pnsp-cl}
  P_{\rm qq}^{\,\rm ns+}(N) &\, = \,& 
  -\:\!\frct{1}{2}\,\* N \*(S-1) \,+\,
  \frct{1}{2} \* \ar \* (2\,\*\cf-\b0) \* (S^{\,-1}-1)
\nn\\ & & \!\mbox{}
  + \frct{1}{96\*\cf}\, \* \ar \* N\*\, \Big\{ \big( [156 - 960\*\,\z2]\,\*\cfs 
 - [80 - 1152\*\,\z2]\,\*\ca\*\cf - 360\*\,\z2\*\cas - 100\,\*\b0\*\cf
\nn\\ && \mbox{\hspp\hspp\hspp}
 + 3\*\bb02 \big) \*(S-1)
 \,+\, 2\,\* \big( [12 - 576\*\,\z2]\,\*\cfs + [40 + 576\*\,\z2]\,\*\ca\*\cf 
 - 180\*\,\z2\*\cas 
\nn\\ && \mbox{\hspp\hspp\hspp}
 + 56\,\*\b0\*\cf - 3\*\bb02 \big) \* (S^{\,-1}-1) 
 \,+\, 3\,\*(2\,\*\cf - \b0)^2\*(S^{\,-3}-1)\Big\} 
\eea
with $S \,=\, ( 1 - 4\:\!\bar{\xi} )^{1/2}$ and 
$\bar{\xi} \,=\, 2\,\cf\ar N^{\,-2} $, of which only the leading-logarithmic
first term was known before \cite{BVns}. While Eq.~(\ref{Pnsp-cl}) is formally
analogous to Eq.~(\ref{PTgg-cl}), the different sign of $\bar{\xi}$ as compared
to $\xi$ leads to a qualitatively different $x$-space behaviour, see 
Ref.~\cite{KV1} where also the resummed coefficient functions $F_{2,\rm ns}$ 
and $F_{L,\rm ns}$ and the flavour-singlet results will be presented.

\section{Summary and Outlook}

\noindent
We have derived all-order results for the highest three (in a few cases four)
large-$x$ and small-$x$ double-logarithmic contributions to spacelike and 
timelike splitting functions and to coefficient functions for inclusive DIS 
and SIA. These results have been obtained from NNLO (in a few cases
effectively N$^3$LO) fixed-order results in a `bottom-up' approach using the 
large-$x$ behaviour of physical evolution kernels, large-$x$ iterations of 
unfactorized structure and fragmentation functions, and KLN-related large-$x$ 
and small-$x$ decompositions of the unfactorized expressions in dimensional 
regularization combined with constraints imposed by the mass-factorization 
relations.

\pagebreak

At least in the form presented above, the last and so far most powerful
approach is applicable neither to inclusive lepton-pair and Higgs production
at large $x$, nor to the odd Mellin-$N$ based structure functions such as
$F_3^{\nu+\bar{\nu}}$ in charge-current DIS and $g_1^{}$ in polarized DIS.
In the former cases, Eq.~(\ref{FThatDec}) holds in the $\x1^{\,-1}$ soft-gluon
limit with $n_0^{} = 0$ and $n_1^{} = 2$ \cite{MMvN,MV1} but, for example, 
additional odd-$k\eta_1$ terms which spoil most of the predictivity are 
found to be required in the $qg$ channel of the Drell-Yan process 
\cite{LPthesis}. Without additional theoretical insight, which may come from 
the more rigorous approaches pursued in Refs.~\cite{LMSW}, it will also not be 
possible to improve upon our present `N$^n$LO implies N$^n$LL' accuracy in 
cases for which Eq.~(\ref{FThatDec}) is applicable.

\vspace*{0.5mm}
Most of the results in 
Refs.~\cite{MV3,MV5,SMVVs5,LL2010,AV2010,ASV,AV2011,Rcor11,KVY,ALPV,KV1}
will not have a direct phenomenological impact, but will hopefully prove 
useful in conjunction with future developments such as, e.g., a computation
of fourth-order moments of structure functions analogous to 
Refs.~\cite{3loopN}. 
The exception are the $\x1^0$ contributions to the non-singlet DIS coefficient 
functions, which should be of interest for high-precision analyses of 
structure functions at large $x$, and the $x^{\,-1}$ small-$x$ timelike 
splitting functions as discussed above. In both cases the relative 
size of the known N$^n$LL contributions indicates that an improvement on the 
present accuracy is required for quantitatively reliable results.

\begin{figure}[t]
\vspace*{-1mm}
\centerline{\hspace*{-2mm}\epsfig{file=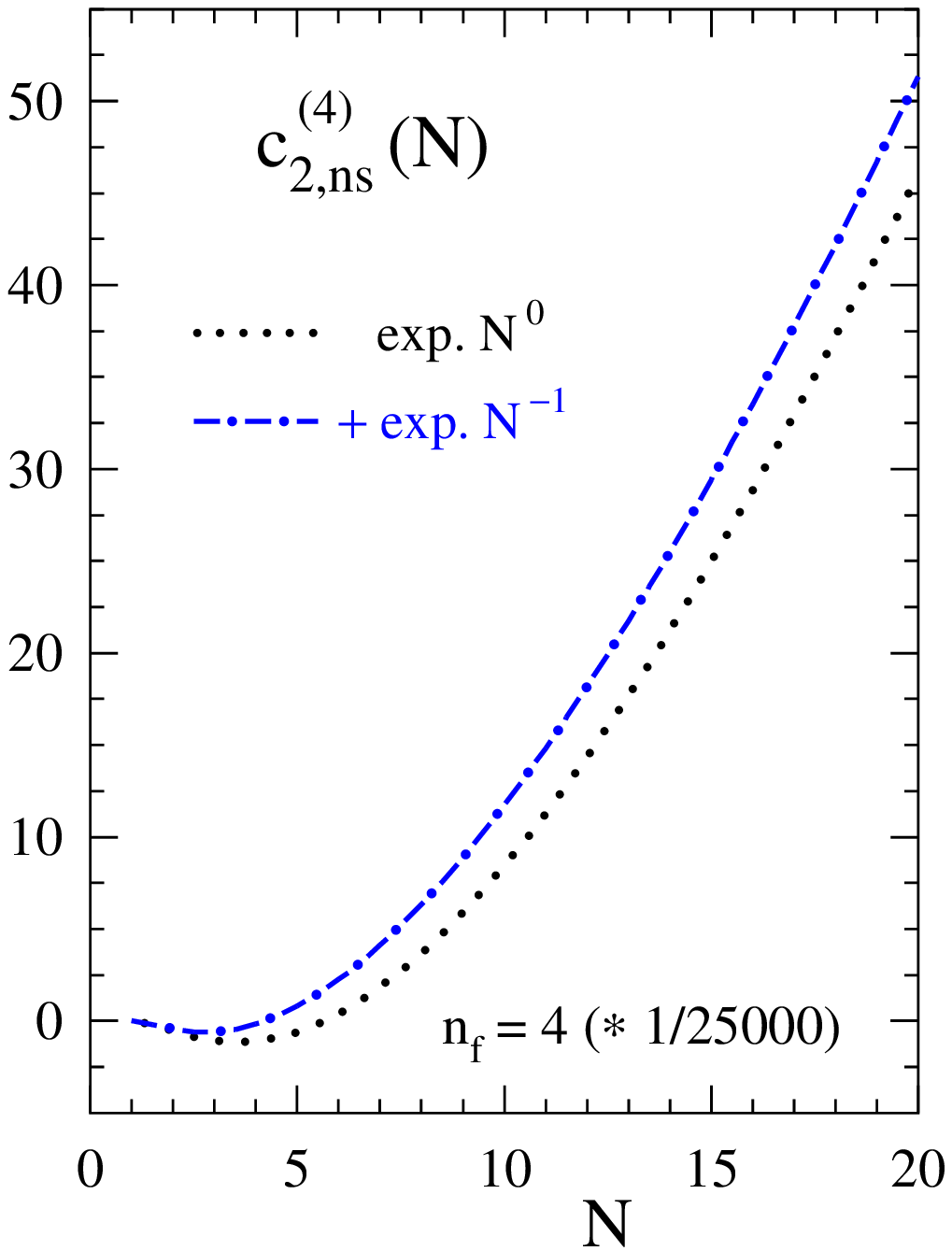,width=6.6cm,angle=0}%
\hspace*{-2mm}\epsfig{file=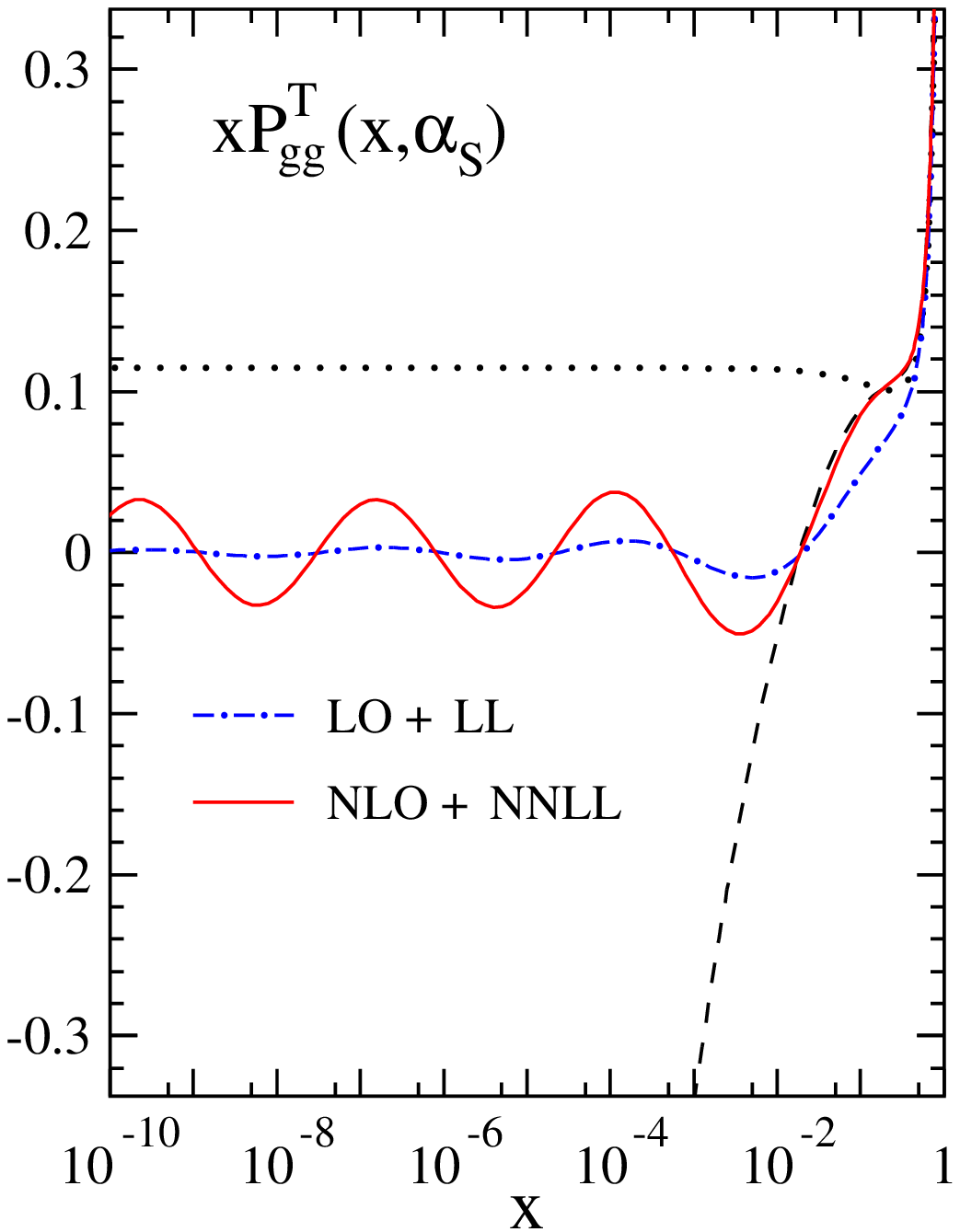,width=6.7cm,angle=0}}
\vspace{-2.5mm}
\caption{ \label{fig2}
 Left: the fourth-order coefficient function for the structure function 
 $F_{2,\rm ns}$. Shown are the large $N$ estimates by the known seven (of 
 eight) $\ln^n N$ soft-gluon contributions and by adding the highest four 
 (of seven) $N^{\,-1}\ln^n N$ corrections.
 Right: the timelike gluon-gluon splitting functions (multiplied by $x$) 
 for a very wide range of the momentum fraction $x$ at a value of $\as$
 corresponding to $Q^{\,2} \simeq M_{Z\!}^{\:2\!}$. The all-$x$ ($N\!=\!1$ 
 finite) LO$\,+\,$LL and NLO$\,+\,$NNLL results are compared to the LO 
 and NLO approximations valid only at large $x$.}
\vspace*{-3mm}
\end{figure}

\vspace*{-1mm}
\subsection*{Acknowledgements}

\noindent
The research summarized here has been supported by
the UK Science \& Technology Facilities Council (STFC) 
and in part by
the European-Union funded network LHCPhenoNet,
the Helmholtz Association of German Research Centres and the German Research
Foundation (DFG),
the Dutch Foundation for Fundamental Research of Matter (FOM),
and 
the Natural Sciences and Engineering Research Council (NSERC) of~Canada.

\end{document}